\documentclass[10pt,conference]{IEEEtran}
\usepackage{graphics,graphicx,cite,subfigure,multicol,picinpar,url,xspace}
\usepackage{times,graphicx,epsfig,epic,eepic,latexsym}
\usepackage{amsfonts,amssymb,amsmath,amsthm,times}
\newtheorem{theorem}{Theorem}
\newtheorem{definition}{Definition}
\newtheorem{lemma}{Lemma}

\newtheorem{claim}{Claim}
\newcommand{\cA}{{\mathcal{A}}}

\title{Network error correction with limited feedback capacity}
\author{\authorblockN{Yanbo Yang$^\ast$, Tracey Ho$^\dagger$ and Wentao Huang$^{\star}$}
\authorblockA{$^\ast$Xidian University, $^\dagger$California Institute of Technology}
Email: $^\ast$ yanboyang@mail.xidian.edu.cn, $^\dagger$tho@caltech.edu and yelo@caltech.edu}

\begin{document}
\maketitle

\begin{abstract}
We consider the problem of characterizing network capacity in the presence of adversarial errors on network links,
focusing in particular on the effect of low-capacity feedback links across network cuts. In~\cite{kim2010necjournal}, 
the authors gave cut-set bound for unequal links capacity network and achievable strategy for coding, decoding and detection
,which only cover the network with large enough feedback across the network, allowing intermediate nodes
reliably transporting part of code by some proper code back to supper source. However when feedback has small capacity,
the strategy will be failed , the intuition is that you cannot have a simple code on low capacity feedback. 
We analysis thoroughly the behavior of adversary on small capacity of feedback link, which turns out that it's more complex 
than large-capacity feedback, and give corresponding strategies. We give a new outer bound as
well as a new achievable strategy, and show a family of networks where the inner and outer bounds coincide.
\end{abstract}

\section{Introduction}
The problem of reliable network communication in the presence of adversarial link errors was first considered by Yeung and
Cai~\cite{cai2006network,yeung2006network} for the case of networks with equal capacity links. They
generalized the Hamming bound, the Singleton bound,
and the Gilbert-Varshamov bound from classical error correction coding to network coding. In this problem, an adversary can arbitrarily corrupt
information on a set of $z$ network links whose locations are unknown to the network user. Yeung and Cai~\cite{cai2006network,yeung2006network} 
showed that the multicast capacity is given by $m-2z$, where $m$ is the minimum source-sink cut
capacity, and that the capacity can be achieved by linear network coding.

However in previous work, the authors assume unit link capacity. In the error-free case, any link $l$ with capacity $r$ can be represented by $r$
edges of capacity one without loss of generality, but in the case with errors, there is a loss of generality in assuming that errors occur
independently on the unit capacity edges.

In recent work~\cite{kim2009network,kim2010nec,kim2010necjournal}, the authors considered the case of networks with
unequal link capacities. They showed that, unlike the case of equal link capacities, feedback across network cuts can increase 
the error correction capacity.  They provided upper bounds on capacity, and coding strategies that achieve the upper bounds in a family 
of zig-zag networks with feedback links of sufficient capacity. The related problem of network error correction with adversarial nodes 
was considered by~\cite{kosut2009nonlinear,kosut-polytope}.

In this paper, we consider zig-zag network with small-capacity feedback links, 
which we have discussed in \cite{Feedbackita} under specific condition of
feedback link, and also for which previous bounds in~\cite{kim2010necjournal} 
are not tight.  We provide a new upper bound and a new coding strategy
that achieves the upper bound in a  family of zig-zag networks with small feedback capacity.

\section{Preliminary}
We consider a communication network represented by a directed acyclic graph
$\mathcal{G} = (\mathcal{V},\mathcal{E})$. Source node $s\in
\mathcal{V}$ transmits information to the sink nodes $u\in
\mathcal{V}$. Let $r(a,b)$ denote the capacity of edge $(a,b)\in\mathcal{E}$. 
We assume that code alphabet $\mathcal{A}$ is $GF(q)$ for some
large enough $q$. We can view  an error vector on a
link $l\in\mathcal{E}$ as set of $r(l)$ symbols in $\cA$, where
the output $y_l$ of link $l$ equals the  mod $q$ sum of the input
$x_l$ to link $l$ and the error $e_l$ applied to link $l$. We say
that there are $\tau$ error links in the network if $e_l\neq 0$ on
$\tau$ links.
\begin{definition}
A network code is $z$-error link-correcting if it can correct any
$\tau$ adversarial links for $\tau\leq z$. That is, if the total
number of adversarial links in the network is at most $z$, then the
source message can be recovered by all the sink nodes
$u\in\mathcal{U}$.
\end{definition}

Let $(A,B)$ be a partition of $\mathcal{V}$, and define the cut for
the partition $(A,B)$ by
\[cut(A,B)=\{(a,b)\in\mathcal{E}:a\in A, b\in B\}.\]
$cut(A,B)$ is called a cut between two nodes $a$ and $b$ if $a\in A$
and $b\in B$. The links in $cut(A,B)$ are called the forward links of the
cut. The links $(a,b)$ for which $a\in B$, $b\in A$ are called the
feedback links of the $cut(A,B)$. The capacity of a cut is the sum
of the capacities of the forward links of the cut.

For any cut $Q=cut(P,\mathcal{V}\backslash P)$, let 
$Q^R$ denote the set of feedback links across cut $Q$. We say that a  feedback link $l\in Q^R$ is
directly downstream of a forward link $l'\in Q$ (and that $l'$ is
directly upstream of $l$) if there is a directed path starting from
$l'$ and ending with $l$ that does not include other links in $Q$ or
$Q^R$. 

\section{New upper bound}
We introduce a new upper bounding approach which considers confusion between two possible sets
of $z$ forward adversarial links, when there exist two codewords that differ in these forward links but coincide 
in the values of their directly downstream feedback links and the remaining forward links. The bound is then the 
sum of the capacities of these feedback links plus the capacities of the remaining forward links. Therefore, this 
bound is useful when the feedback link capacities are sufficiently small.

To state this result formally, consider
any cut $Q=cut(P,\mathcal{V}\backslash P)$,
and two disjoint sets of forward links
$Z_1,Z_2\subset Q$ where  $|Z_i|\leq z$ for $i=1,2$. Let $W_1$  be the set of links in $Q^R$ which are directly
downstream of a link in $Z_1$ and upstream of a link in
$Q\backslash Z_1$. Let $W_2$  be the set of links in $Q^R$ which
are directly downstream of a link in $Z_2$ and upstream of a link in
$Q\backslash Z_2$. 

\begin{theorem}\label{bound2}
Let $$M=\sum_{(a,b)\in ((Q\backslash Z_1)\backslash
Z_2)\cup W_1\cup W_2}r(a,b)$$ denote the sum of capacities of forward links in $(Q\backslash$
$Z_1)\backslash Z_2$ and feedback links in $W_1$ and
$W_2$. If no link in $Z_2$ is directly
upstream of any link in $W_1$, and no link in $Z_1$ is directly
upstream of any link in $W_2$, then the capacity is at most $M$.
\end{theorem}

\begin{IEEEproof}
We assume that the codebook $X$ contains more than $ q^{M}$ codewords, and show that this leads to a
contradiction. Let $k$ denote the number of links on the cut $Q$, and let $m=|Z_1|$, $n=|Z_2|$. 

Since $|X|>q^{M}$, from the definition of $M$, there exist two
distinct codewords $x,x'\in X$ such that the error-free outputs on the
links in $(Q\backslash Z_1)\backslash Z_2$ and $W_1\cup W_2$ are the
same. 
The corresponding observations on the sink side of the cut are
\begin{eqnarray*}O(x) & = &\{y_1,..,y_{k-m-n},u_{1},..,u_{m},w_1,..,w_{n}\}\\
O(x') & = &\{y_1,..,y_{k-m-n},u_{1}',..,u_{m}',w_1',..,w_{n}'\},\end{eqnarray*}
where  $(y_1,..,y_{k-m-n})$ denote the error-free outputs on the
links in $(Q\backslash Z_1)\backslash Z_2$ for $x$ and $x'$;
$(u_{1},..,u_{m})$ and  $(u_{1}',..,u_{m}')$ denote the error-free
outputs on the links in $Z_1$ for $x$ and $x'$ respectively; and
$(w_{1},..,w_{n})$ and $(w_{1}',..,w_{n}')$ denote the error-free
outputs on the links in $Z_2$ for $x$ and $x'$ respectively.

Since no link in $Z_2$ is directly
upstream of any link in $W_1$, the values on $W_1$ are determined
by the values on $Q\backslash Z_2$. Similarly, since  no link in $Z_1$ is directly
upstream of any link in $W_2$, the values on
$W_2$ are determined by $Q\backslash Z_1$.

We will show that it is possible for the adversary to produce
exactly the same outputs on all the channels in $Q$ under  $x$ and
$x'$ when errors occur on at most $z$ links. 

Assume the input of network is $x$. The adversary could choose forward links set $Z_1$ as its $z$ adversarial
links, and apply errors on $Z_1$ to change the
output from $u_{i}$ to $u_{i}'$ $\forall 1\leq i\leq m$. Note that
the values on $W_1$ are determined by the values on $Q\backslash Z_2$, which, under these errors, are same as for $x'$,  and that
the values on $W_1$ are the same for $x$
and $x'$.  Therefore, the
values on $W_1$ are not changed, and thus the values on $Q\backslash
Z_1$ are not affected. The observations on the sink side of the cut are
$\{y_1,..,y_{k-m-n},u_{1}',..,u_m',w_1,..,w_n\}$.

When codeword $x'$ is transmitted, the adversary could choose forward
links set $Z_2$ as its $z$ adversarial links, and apply
errors on them to change $(w_1',..,w_n')$ to $(w_1,..,w_n)$. Similarly, since 
the values on $W_2$ are determined by
the values on $Q\backslash Z_1$, which, under these errors,  are  the same as for $x$, 
and since the values on $W_1$ are the same for $x$
and $x'$, the values on $W_2$ and on $Q\backslash
Z_2$ are not affected. The observations on the sink side of the cut are 
$\{y_1,..,y_{k-m-n},u_{1}',..,u_m',w_1,..,w_n\}$, the same output as before.
\end{IEEEproof}
A number of variations of this result are possible.  For instance, if there are no links in $Q^R$ which
are directly downstream of a link in $Z_2$ and upstream of a link in
$Q\backslash Z_2$   (i.e.~$W_2$ is empty), we can redefine $W_1$ to be the set of links in $Q^R$ which are directly
downstream of a link in $Z_1$ and upstream of a link in
$(Q\backslash Z_1)\backslash Z_2$, which is smaller compared to  the previous definition.

\begin{figure}[h]
\begin{center}
  \includegraphics[scale=0.8]{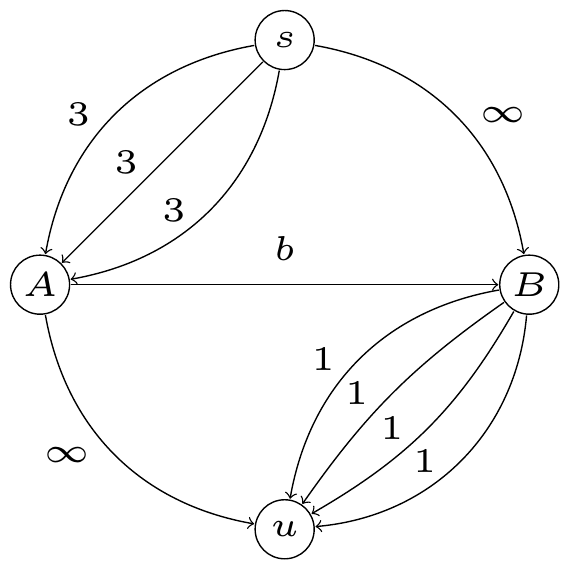}
\end{center}
\caption{Four node acyclic networks:Given the cut $Q=cut(\{s,B\},\{A,t\})$, 
unbounded reliable communication is allowed from source $s$ to its
  neighbor $B$ on one side of the cut and from node $A$ to sink $u$ on the 
other side of cut, respectively. There is a feedback link from $A$ to $B$
with capacity $b$}
\label{figure:four}
\end{figure}

For the example network in Fig.~\ref{figure:four}, with two adversarial links, our previous result 
in~\cite{kim2010necjournal} gives a bound of $8$,
whereas Theorem 1 gives a bound of $5+b$, which is tighter when $b <3$.

\section{Preliminary for achieve strategy }
\label{s-condition}

\begin{figure}[h]
\begin{center}
  \includegraphics[scale=0.8]{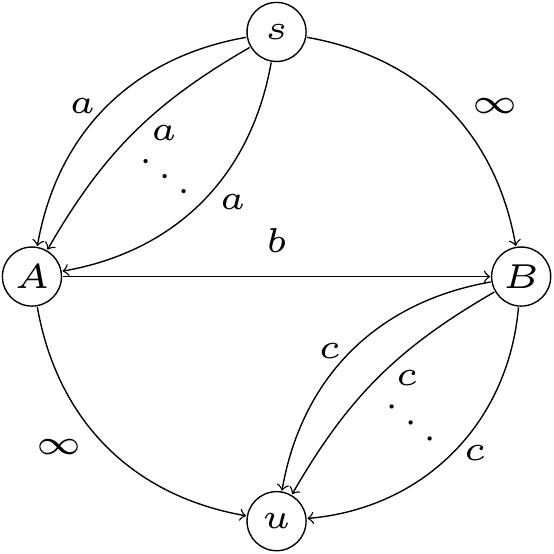}
\end{center}
\caption{Four nodes acyclic network: $n$ forward links from node $s$ to 
intermediate node $A$, $m$ links between node $B$ and sink $u$
, and two unbounded links.}
\label{fan}
\end{figure}

We define a family of zig-zag acyclic network model first. Denote Fig.~\ref{fan} by $\mathcal{G}$,
let $s$ is the source node and $u$ is the sink, node $A$ and $B$ are two intermediate nodes.  
let $r(\cdot)$ represent the capacity of link, given $r(s,A)=a,r(B,u)=c, r(A,B)=b$ and $a>c, a>b$. Also 
unbounded reliable communication is allowed from source $s$ to $B$ and from node $A$ to sink $u$.
Let $n$ be the number of links from source node to node $A$ and $m$ be the number of links from node $B$ to the sink.
We will call upstream for links between source and node $A$, downstream for links between intermediate node $B$ and sink. 

By Theorem~1, we focus on the family of zig-zag network, the upper bound of which is
$$
UB = (n-z)a+(m-z)c+b
$$

According to \cite{kim2010necjournal}, the tight bound of zig-zag network is strongly related with
topology of network, in order to provide a proper coding strategy, we have to first discuss
properties of topology of network.

To find these conditions, we compare $UB$ with cut set bound considered in \cite{kim2010necjournal}. 
Strictly, there are three variations of cut set bound according to topology of network in \cite{kim2010necjournal}.
We describe these possible bounds  briefly as following:
\begin{enumerate}
 \item {\bf $SB1$}: Adversary attacks $z_1$ downstream links and no more than $2z-z_1$ other downstream links on two codewords respectively, 
 which they implicate each other.
 \item {\bf $SB2$}: Adversary erases feedback link and we bound the remaining network by Singleton bound .
 \item {\bf $SB3$}: Adversary attacks feedback link and $z_1$ upstream links implicating $z_2$ attack on downstream, and attacks 
 $z_2$ on downstream links implicating $z_1$ attack on downstream and feedback, $z_1+z_2\le 2z-1$.
\item {\bf $SB4(UB)$}: Adversary attacks $z_1$ upstream links and $z_2$ downstream links to implicate 
     each other with the same output on feedback link, $z_1+z_2\le 2z$.
\end{enumerate}

We categorize zig-zag network into four categories by observation of number of upstream and
downstream,

{\bf Category 1.} $n\ge 2(z-1), m\ge 2z$

According to above describe about $SB1$ and $SB2$, we give
\begin{eqnarray}
SB1 &=& na+(m-2z)c \nonumber \\
SB2 &=& (n-2(z-1))a+mc  \nonumber
\end{eqnarray}
Suppose $UB$ is tighter than $SB1$ and $SB2$, then 
\begin{eqnarray}
0 &<& SB1-UB  \nonumber \\
0 &<& na+(m-2z)c -((n-z)a+(m-z)c+b )  \nonumber \\
b &<& z(a-c) \nonumber
\end{eqnarray}
and
\begin{eqnarray}
 0 &<& SB2-UB \nonumber \\
 0 &<& (n-2(z-1))a+mc-((n-z)a+(m-z)c+b) \nonumber \\
 b &<& zc-(z-2)a \nonumber
\end{eqnarray}

Thus we have,
\begin{eqnarray}
b < \min \{ z(a-c), zc-(z-2)a \} \label{con1}
\end{eqnarray}
and new bound is tightest in network $\mathcal{G}$ under $(\ref{con1})$.

{\bf Category 2.} $z\le n < 2(z-1), m\ge 2z$

We have 
\begin{eqnarray}
SB1 &=& na+(m-2z)c \nonumber \\
SB2 &=& (m-(2(z-1)-n))c \nonumber
\end{eqnarray}
Since $UB< SB1, UB< SB2$, it's not hard to have 
\begin{eqnarray}
b<\min\{ z(a-c), (n-z+2)c-(n-z)a \} \label{con2}
\end{eqnarray}

Clearly, $(\ref{con1})$ and $(\ref{con2})$ exactly define a family of networks in which new bound we proposed is tight.

For $SB3$, it is a trivial case that $UB<SB$, since
\begin{eqnarray}
 SB3 &=&(n-(z-1))a+(m-z)c \nonumber \\
     &=& (n-z)a+(m-z)c+a  \nonumber
\end{eqnarray}
where $b<a$ by definition of four-node network.

{\bf Category 3.} $n\ge 2(z-1), m < 2z$

According to above describe on possible bounds, we have 
\begin{eqnarray}
 SB1 &=& (n-(2z-m))a \nonumber \\
 SB2 &=& (n-2(z-1))a+mc \nonumber \\
     &=& (n-z)a+(m-z)c+zc-(z-2)a \nonumber \\
 SB3 &=& (n-(z-1))a+(m-z)c \nonumber \\
     &=& (n-z)a+(m-z)c+a \nonumber 
\end{eqnarray}

Suppose $UB$ is tighter than above bounds, compare it with $SB1,SB2,SB3$, 
not hard to have the following inequalities,
\begin{eqnarray*}
b &<& zc-(z-2)a  \\
b &<& (m-z)(a-c) \\
b &<& a         
\end{eqnarray*} 
which follows $UB<SB1, UB<SB2, UB<SB3$ respectively,
thus we have
\begin{eqnarray}
b<\min \{ zc-(z-2)a,(m-z)(a-c) \} \label{con3}
\end{eqnarray}

{\bf Category 4.} : $z\le n < 2(z-1), m < 2z$

Similarly, we have $SB1,SB2,SB3$, 
\begin{eqnarray}
 SB1 &=& (n-(2z-m))a \nonumber \\
     &=& (n-z)a+(m-z)c+(m-z)(a-c) \nonumber \\
 SB2 &=& (m-(2(z-1)-n))c \nonumber \\
     &=& (n-z)a+(m-z)c+(n-z+2)c-(n-z)a \nonumber \\
 SB3 &=& (n-(z-1))a+(m-z)c \nonumber \\     
     &=& (n-z)a+(m-z)c+a \nonumber 
\end{eqnarray}
Suppose $UB$ is tighter than the other bounds, we have the tight condition,
\begin{eqnarray*}
b &<& (n-z+2)c-(n-z)a \\
b &<& (m-z)(a-c)       \\
b &<& a              
\end{eqnarray*}
which gives
\begin{eqnarray}
b < \min\{ (n-z+2)c-(n-z)a, (m-z)(a-c) \} \label{con4}
\end{eqnarray}

Note that in this paper, we will call $(\ref{con1}-\ref{con4})$ as {\bf tight conditions},
We formalize the upper bound to be 

\section{Encode}
\label{encode}

Let $\{\mathcal{X},\mathcal{Y}\}$ be message set and $k_1=n-z,k_2=\lceil \frac{b}{a-c} \rceil$, 
\begin{eqnarray*}
\mathcal{X}   &=& \{ x_{i,j}| 1\le i \le a-c, 1\le j \le (n-z)+k_2 \} \\ 
\mathcal{Y}   &=& \{ y_{i,j}| 1\le i\le c, 1\le j\le m+n-2z  \} 
\end{eqnarray*} 

{\bf Notice} that for $\mathcal{X}$, we define only when $\frac{b}{a-c}$ is integer, equality holds for
$j$, otherwise $1\le j < (n-z)+k_2$ such that $|\mathcal{X}|=UB-|\mathcal{Y}|$.

Each message symbol is i.i.d over $GF(q)$ and $q$ is a prime power. Let  
$X$ be $(a-c)\times n$ matrix, $i$th row of which is defined as
\begin{eqnarray*}
 (r_{i,1},r_{i,2},\ldots,r_{i,n}) = (x_{i,1},x_{i,2},\ldots,x_{i,k_1+k'_{i,2}})G 
\end{eqnarray*}
where $x_{i,j}\in \mathcal{X}$ and $\sum_{i=1}^{a-c}k'_{i,2}=b,0\le k'_{i,2} \le k_2$
\begin{eqnarray}
G = (I_{(k_1+k'_{i,2})\times (k_1+k'_{i,2})} \;\; E)\label{equation:G}
\end{eqnarray} 
where $G$ is generator matrix for a $(n,n-k_1-k'_{i,2})$ systematic MDS code, we denote entries
of $E$ by $\eta_{i,j},i=1,\ldots,k_1+k'_{i,2},j=1,\ldots,n-k_1-k'_{i,2}$, which are chosen uniformly at random
from nonzero entries of a large finite field. 

Not hard to check that we can choose $k'_{i,2}\in [0,k_2]$ for each row of $X$ properly to encode
$|\mathcal{X}|=UB-|\mathcal{Y}|$ message symbols.

Define $Y$ is a $c\times (n+m-2z)$ matrix, the $(i,j)$th entry of matrix $Y$ is given by
$y_{i,j}\in\mathcal{Y}, 1\le i\le c, 1\le j\le n+m-2z$, and define $L$ is a $c\times 2z$ matrix, 
the $(i,j)$th entry of matrix $L$ is denoted by $l_{i,j}, 1\le i\le c, 1\le j\le 2z$,
which is linear combination of all the message symbols $\{\mathcal{X},\mathcal{Y}\}$, 
in that any subset of symbols is maximally independent. 
Note that there is large enough $q$ such that any subset of symbols in $Y$ are
maximally independent over $GF(q)$.

On network $\mathcal{G}$, we define codeword for given message sequence $(\mathcal{X,Y})$ as follows,
\begin{eqnarray}
C=\left (
\begin{matrix}
X   & \mathbf{0} \\
Y   & L
\end{matrix}
\right ) \nonumber
\end{eqnarray} 
where $\mathbf{0}$ is zero matrix, the dimension of which is $(a-c)\times m$. Then the dimension of $C$ is $a\times (m+n)$.
Source $S$ and node $B$ cooperate to send codeword $C$. Source $S$ sends the first columns
$i=1,\ldots,n$ of $C$ to $A$ and node $B$ sends the columns $i=n+1,\ldots, n+m$ to sink.

If no errors have previously been detected, the signal sent by node $A$ on the feedback link is
defined as follows. Consider arbitrary row of $X$, we rewrite the row vector as following for easily analysis, 
and we ignore the index of row for simplicity. 
\begin{eqnarray}
&& (r_1\ldots r_{k_1},r_{k_1+1}\ldots r_{k_1+k'_2},r_{k_1+k_2'+1}\ldots r_{n}) \nonumber \\
&=&(x_1\ldots x_{k_1}, v_{1}\ldots v_{k'_2},r_1\ldots r_{k_3}) \label{subx1}
\end{eqnarray}
where $k_3=n-k_1-k_2'$ and let
$ {\bf x}=\{ x_i=r_i, i=1,\ldots, k_1 \}, {\bf v}=\{ v_i=r_{k_1+i}, i=1,\ldots, k_2' \},
{\bf r} = \{ r_i, i=1,\ldots, n-k_1-k_2'\}$
where from (\ref{equation:G}), $r_j=\sum_{i=1}^{k_1+k_2'}\eta_{i,j}x_{i},j=1,\ldots,n-k_1-k_2' $ and
$k_2'$ represents $k_{i,2}'$, $\eta_{i,j}$ represents the $(i,j)$ elements of $E$.

Now consider a received row vector of $\hat{X}$ at node $A$,
\begin{eqnarray}
  (\hat{x}_1\ldots\hat{x}_{k_1},\hat{v}_{k_1+1} \ldots \hat{v}_{k_1+k_2'},\hat{r}_{1}\ldots \hat{r}_{k_3}).
\end{eqnarray}where from (Node $A$) sends the feedback symbols 
\begin{eqnarray}
\{\delta_i =  \hat{v}_i-(\eta_{1,i}\hat{x}_1+\ldots+\eta_{k_1,i}\hat{x}_{k_1}) | i=1,\ldots,k_2' \} 
\end{eqnarray} 
corresponding to each row of $\hat{X}$.

Then we define {\em claim} message on node $B$ which will be used when new erroneous is detected on upstream.
Let {\em claim} message $W$ be $(a-c)\times m$ matrix such that $(X,W)$ is an $(n+m,n+m-k_1-k_2)$ MDS code.

\section{Encoding Error Signaling}
\label{EES}

We will describe strategy to show how intermediate node $A$ and $B$ cooperate to
detect adversary on upstream in this section. 

Strategy will show that either intermediate node $A$ or $B$ could find out  
when erroneous is injected first time on upstream,  
and Lemma~\ref{Singleton-new} in following shows that if sink has identified two adversary
links successfully on upstream, sink can decode without feedback and 
any extra information from intermediate nodes, thus we will focus on the following 
{\bf scenario} of adversary behavior,
\begin{enumerate}
\item Adversary only attacks forward link in first attack.
\item Adversary attacks the other forward links to hide the changing of attack behavior in subsequent attack.
\end{enumerate}

\begin{lemma}
Given two adversary links from source $S$ to node $A$ are identified, the sink can decode correctly
with remaining codeword without information from feedback link.
\label{Singleton-new}
\end{lemma}
\begin{IEEEproof}
Suppose the number of upstream links identified by sink is $x$, denote Singleton bound on remaining network by $SB$.

{\bf Category 1.} $n>2(z-1)$

Singleton bound, $SB$, on the remaining network is 
\begin{eqnarray}
SB &=& ( n-x-2(z-x))a+mc \label{sb1} \\
   &=& ( n+x-2z)a + mc \nonumber 
\end{eqnarray}
Then define
\begin{eqnarray}
f(x) &=& SB-UB \nonumber \\
     &=& (n+x-2z)a+mc-((n-z)a+(m-z)c+b) \nonumber \\
     &=& ax-z(a-c)-b \nonumber
\end{eqnarray}

Observe zero point of $f(x)$, we have
\begin{eqnarray}
x &=& \frac{z(a-c)+b}{a} \label{eq11} \nonumber \\
  &<&\frac{z(a-c)+zc-(z-2)a}{a} \\
  &=& 2 \nonumber 
\end{eqnarray}
where $(\ref{eq11})$ follows $(\ref{con1}),(\ref{con3})$ . 

Then $f(x)$ is a strictly monotonous increase function for given network. 

{\bf Category 2.}  $z\le n \le 2(z-1)$

Consider the $SB$ on remaining network,
\begin{eqnarray}
SB = \left \{
\begin{matrix}
 (n+x-2z)a+mc &\; x > 2z-n \\
 (n+m+x-2z)c  &\; x \le 2z-n
\end{matrix}
\right. \nonumber 
\end{eqnarray}

Then we have $f(x)$ denoted as
\begin{eqnarray}
 f(x) &=& SB-UB \nonumber \\
      &=& \left \{
\begin{matrix}
 ax-z(a-c)-b  &\; x > 2z-n \\
 xc-(n-z)(a-c)-b     &\; x \le 2z-n 
\end{matrix}
\right. \nonumber
\end{eqnarray}
Consider zero point of $f(x)$ when $x\le 2z-n$,
\begin{eqnarray}
 0 &=& f(x) \\
 0 &=& xc-(n-z)(a-c)-b\nonumber \\
x &=& \frac{(n-z)(a-c)+b}{c}  \nonumber \\
  &<& \frac{(n-z)(a-c)+(n-z+2)c-(n-z)a}{c} \label{eq222}\\
  &=& 2 \nonumber 
\end{eqnarray}
$(\ref{eq222})$ follows $(\ref{con2}),(\ref{con4})$. Clearly, $f(x)$ is a strictly monotonous increase function when $x\le 2z-n$.
Then we observe $f(x)$ for $x > 2z-n$,
\begin{eqnarray}
 f(x) &=& ax-z(a-c)-b \nonumber \\
      &>& ax-z(a-c)-((n-z+2)c-(n-z)a) \label{eq333} \\
      &>& ax-z(a-c)-(zc-(n-z)a) \label{eqm4} \\
      &=& (x-2z+n)a  \nonumber  \\
      &>& 0  \label{eq5}
\end{eqnarray}
where $(\ref{eq333})$ follows $(\ref{con2}),(\ref{con4})$, $(\ref{eqm4})$ follows $n\le 2(z-1)$ and $(\ref{eq5})$ follows $x>2z-n$.

Then $f(x)$ is a a strictly monotonous increase function and it has an unique zero point $x=2$. Thus we have $f(x)=SB-UB>0$
when $x=2$, which indicates that Singleton bound on the remaining network is greater
than $UB$ on the original network after the sink identifies $2$ adversary links between source node
and node $A$.

\end{IEEEproof}

\subsection{Preliminary for error signaling} 

For each codeword $C$, intermediate node $A$ receives $\hat{X}$ and $\hat{Y}$. Node $A$ 
checks each row of $\hat{X}$, then according to results of observation, it will send side
information to node $B$, which lets node $B$ have some knowledge of upstream. 

Before error signaling description, we give some preliminary first and 
we assume that feedback link is clean in section~\ref{EES}. Consider a row vector of $\hat{X}$ at node $A$ by
ignoring index of row,
\begin{eqnarray*}
  (\hat{x}_1\ldots\hat{x}_{k_1},\hat{v}_1\ldots\hat{v}_{k_2'},\hat{r}_{1}\ldots\hat{r}_{k_3}) 
\end{eqnarray*}
We define $\Delta$ for each received row vector, 
\begin{eqnarray}
\Delta = \{ \delta_j=\hat{r}_j-\sum_{i=1}^{k_1}\eta_{i,j}\hat{x}_{i}-\sum_{i=k_1+1}^{k_1+k_2'}\eta_{i,j}\hat{v}_{i})  \} \label{Delta2}
\end{eqnarray}
where $j=1,\ldots,k_3$.

The intuition of detection is that intermediate node $A$ and $B$ choose 
proper strategy by observing $(\ref{Delta2})$ or variation of $(\ref{Delta2})$ 
to detect new happened adversary links.

Define claim-sending signal from node $A$, calling it $CS$ for short, once node $B$ 
receives $CS$, it will sends $W$ to sink, it will let sink have $(\hat{X},\hat{W})$, 
which is an MDS code against at most $z$ error links.

\subsection{Error Signal Encoding Analysis}

In this subsection, we will discuss several cases to show that there 
is always a proper strategy for intermediate node $A$ and
$B$ such that:
\begin{itemize}
\item Once adversary conduct the first attack on $X$, either node $A$ or node $B$ has ability to find it.
\item Adversary cannot hide the new error links by corrupt multi-links given there was only one link 
attacked in previous transmission.
\end{itemize}

We simply describe feedback operation for error signal first. For each received row of $\hat{X}$, node
$A$ observes $\Delta$, since $\delta_i\in\Delta$ is derived from redundant symbol on
corresponding row of $\hat{X}$, it can reflect whether $({\bf x},{\bf r})$ is attacked, then based
on observation of $\Delta$, node $A$ will send different error signal messages through feedback.

Also the signals on feedback could be different even for the same observation of $\Delta$ since the
same observations of $\Delta$ from different transmission could be caused by different adversary
behaviors. Thus error signal on feedback that node $A$ sends is formulated by considering the
observation of previous $\Delta$ too. 

And we define a general operation function on feedback,
\begin{eqnarray}
F = \{ \tilde{f}=H(\delta_i,i=1,\ldots,k_2) \},
\end{eqnarray}
where $H(\cdot)$ is a linear function determined by the observation of $\Delta$.

We organize error signaling analysis according to size of $\Delta$ since different
size of $\Delta$ indicates different detection ability on node $A$, and the strategy
used by node $A$ to cooperate with node $B$ have to be different.

Recall definition of $\Delta$, the size of $\Delta$ on each row of $X$ is determined
by $k'_2$. Since $k_2=\lceil \frac{b}{a-c}\rceil$ and $b<z(a-c)$, we have $k_2\le z$,
which says $0\le k_2' \le z$. Given an $X$, different rows can has different size of $\Delta$, 
which are
\begin{enumerate}
\item $|\Delta| = z$ with $k_2'=0$,
\item $2\le |\Delta|< z$ with $0< k_2' \le z-2$,
\item $|\Delta|=0$ with $k_2'=z$.
\item $|\Delta|=1$ with $k_2'=z-1$,
\end{enumerate}

We will discuss them in four cases, in each of which we focus on node $A$ and node $B$
cooperates when erroneous happened first time and node $A$ and node $B$'s behavior 
when new attack happens in subsequent transmission.

{\bf Case-1}: $|\Delta|=z,k_2'=0$

By the condition, the row has structure as
\begin{eqnarray}
(x_1,\ldots,x_{k_1},r_1,\ldots,r_z) \label{received:1}
\end{eqnarray}
where $x_i\in\mathcal{X},i=1,..,k_1$ and $r_j=\sum_{i=1}^{k_1}\eta_{i,j}x_i,j=1,\ldots,z$.

Given a received $(\ref{received:1})$, we have
$$
\Delta=\{\delta_j=\hat{r}_j-\sum_{i=1}^{k_1}\eta_{i,j}\hat{x}_i| j=1,\ldots,z\}
$$
and $\hat{r}_j=\sum_{i=1}^{k_1}\eta_{i,j}x_i+\triangle r_j$ where $x_i$ is clean symbol and $\triangle
r_j$ is injected error symbol, let $\triangle x_i=x_i-\hat{x}_i$, we have
\begin{eqnarray}
\Delta = \{ \delta_j = \sum_{i=1}^{k_1}\eta_{i,j}\triangle x_i+\triangle r_j, j=1,\ldots,z \}
\end{eqnarray}
Let $e$ be number of happened error links, then we consider:

Case-1-1: $e=1$ in first attack.

There are two choices for adversary to conduct the corruption,
\begin{enumerate}
\item[1.1)] Attack one of $x_i, i=1,\ldots,k_1$,
\item[1.2)] Attack one of $r_j, j=1,\ldots, z$.
\end{enumerate}

For (1.1), node $A$ observes all $\delta_j\not=0\in\Delta$ and not hard to see that
it can identify the attacked link.

For (1.2), node $A$ observes single $\delta_i\not=0$ of $\Delta$, and the other $\delta_j$ 
in $\Delta$ equals to zero.

Both of these observation happened in first time, node $A$ sends $CS$ signal to node $B$,
which triggers node $B$ sending $W$ to sink.

Case-1-2: $e\ge 2$ in subsequent attack.

We show node $A$ can detect if there is new error link happened by observe $\Delta$ or a
variation of $\Delta$. In subsequent attack, in order to avoid detecting by node $A$,
adversary should create the same observation of $\Delta$ for node $A$ by attacking new links. 

For (1.1) happened in first attack, we give,
\begin{claim}
\label{claim:1}
 Given single $\hat{x}_i,i\in[1,k_1]$ is attacked in first attack, adversary cannot hide new error link in next attack.
\end{claim}

\begin{IEEEproof}
Assume adversary attacks new links in $X$ in next attack, node $A$ observes the following equations
\begin{eqnarray}
&&\delta_j - \frac{\eta_{i,j}}{\eta_{i,1}}\delta_1  \nonumber \\
&=& \sum_{g=1}^{k_1}\eta_{g,j}\triangle x_g+\triangle r_j
-\frac{\eta_{i,j}}{\eta_{i,1}}(\sum_{g=1}^{k_1}\eta_{g,1}\triangle x_g+\triangle r_1) \nonumber \\
&=& \sum_{g=1}^{k_1\backslash
  i}(\eta_{g,j}-\frac{\eta_{i,j}}{\eta_{i,1}}\eta_{g,1})\triangle x_g +\triangle r_j-\frac{\eta_{i,j}}{\eta_{i,1}}\triangle r_1 \label{GG:1} \\
  && \;\;\;\;\;\;\;\;\;\;\;\;\;\;\;\;\;\;\;\;\;\;\;\;\;\;\;\;\;\;\;\;\;\;\;\;\;\;\;\;\;\;\;\;\;\;\;\ j=2,\ldots,z \nonumber
\end{eqnarray}
In $(\ref{GG:1})$, the term $\triangle x_i$ is canceled ($\triangle x_i$ is on the previous
identified link). Except $\triangle x_i$, there are at most $z-1$ non-zero injected error terms
including $\triangle x_j,\triangle r_j$. Note that we have $z-1$ equations, by property of linear
there must be at least one non-zero equation if new error link happened.
\end{IEEEproof}

For (1.2) happened in first time, node $A$ observed one $\delta_j\not=0$ and all the other $\delta_i=0$ 
in $\Delta$. In subsequent attack, adversary has $z-1$ extra ability to 
corrupt new links, adversary has two
possible attack behavior:
\begin{enumerate}
\item Only attack new single link, i.e., attack either one of ${\bf x}$ or 
$r_l,l\in [1,z]$ such that $\delta_l\not=0, l\not=j$.
\item Attack multi links.
\end{enumerate}
Consider only new single error link happened. If it's happened on one of
$x_i,i=1,..,k_1$, it will cause all $\delta_i\not=0$ in $\Delta$, which will let node $A$ 
know there's new error link happened. If adversary only attacks one of $r_l,l\not=j$, node
$A$ will observe $\delta_l\not=0$ and the other $\delta_j=0$ in $\Delta$, which can trigger node $A$ sending $CS$.

Consider multi error links in next attack. If error only injected in ${\bf r}$, then node $A$ will
observe more than one non-zero $\delta_j$ of $\Delta$; if there is error on ${\bf x}$, assuming
$\tilde{z}$ errors happened on $x_i,i=1,..,k_1$, in order to keep $(z-1)$ $\delta_i=0, i\not=j$, 
adversary needs to attack $z$ extra links since the property of MDS code, however adversary 
only has $z-1-\tilde{z}$ extra ability, it indicates that if any new error link 
happened, node $A$ will observe more than one $\delta_j\not=0$ in $\Delta$. And it means that 
adversary cannot hide multi-error in next attack, i.e., adversary cannot attack multi links such 
that only $\delta_j\not= 0$ and all the other $\delta_i=0$ in $\Delta$.

{\bf Case-2}: $2\le |\Delta| < z, 0<k_2'\le z-2$

By condition, the row has structure
\begin{eqnarray}
(x_1,\ldots,x_{k_1},v_{1},\ldots,v_{k_2'},r_1,\ldots,r_{k_3}) \label{received:2}
\end{eqnarray}
where $2\le k_3\le z$. Given a received $(\ref{received:2})$, we have
$$
\Delta = \{ \delta_j=\hat{r}_j-\sum_{i=1}^{k_1}\eta_{i,j}\hat{x}_{i}-\sum_{i=k_1+1}^{k_1+k_2'}\eta_{i,j}\hat{v}_{i}|j=1,\ldots,k_3\}
$$
where for $j=1,\ldots,k_3$,
\begin{eqnarray*}
\delta_j &=& r_j+\triangle r_j-(\sum_{i=1}^{k_1}\eta_{i,j}\hat{x}_i+\sum_{i=k_1+1}^{k_1+k_2'}\eta_{i,j}\hat{v}_i) \\ 
         &=& \sum_{i=1}^{k_1+k_2'}\eta_{i,j}\triangle x_i+\triangle r_j
\end{eqnarray*}
where $\triangle x_i=x_i-\hat{x}_i$ and $\triangle x_i,\triangle r_i$ are injected error symbol.

First we show node adversary cannot hide erroneous on both of node $A$ and $B$ at the same time,
which implies that once error happened first time either node $A$ or node $B$ can detect it.

Assume upstream link/links was/were attacked, the only way to avoid detection on node $A$ is given observation
of all $\delta_j=0\in\Delta$, which is
$$\sum_{i=1}^{k_1+k_2'}\eta_{i,j}\triangle x_i+\triangle r_j=0,j=1,\ldots,k_3$$
Without error detection, node $A$ sends set $F$ to node $B$, where
\begin{eqnarray}
F=\{f_j=\hat{v}_j+\sum_{i=1}^{k_1}\theta_{i,j}\hat{x}_{i}|j=1,\ldots,k_2'\} \label{feedback:1}
\end{eqnarray}
where $\theta_{i,j}\in H$ and $H$ is $k_1\times k_2'$ matrix, each entry of which is chosen uniformly at
random from nonzero entries of a large finite field such that columns have maximal independence.
When node $B$ receives set $F$, recall that it knows all sending message by source, it observes
for $f_j\in F,j=1,\ldots,k_2'$.
\begin{eqnarray}
\omega_{j} &=& f_j-v_j-\sum_{i=1}^{k_1}\theta_{i,j}x_i \label{omega:1}\\
          &=& (\hat{v}_j-v_j)+\sum_{i=1}^{k_1}\theta_{i,j}(\hat{x}_i-x_i) \nonumber \\
          &=& \triangle v_j+\sum_{i=1}^{k_1}\theta_{i,j}\triangle x_i \nonumber
\end{eqnarray}
since $\triangle v_j,\triangle x_j,\triangle r_j$ are injected error symbols,
and adversary only can affect $z$ links at most. Also we have $k_3+k_2'=z$,
therefor there must be at least one $\omega_i\not=0$, otherwise it implies
all injected $\triangle v_j,\triangle x_j,\triangle r_j$ equal to zero, which
contradicts assumption.

So once upstream links are attacked first time, either node $A$ or node $B$ 
can detect it, and if there are more than one upstream links attacked, by Lemma~1,
it won't need detection of intermediate nodes in subsequent transmissions. 
Then we consider the attacking scenario as

Case-2-1: $e=1$ in first attack.

Consider a received vector as $(\ref{received:2})$ with a single corrupted symbol,
the possible position is
\begin{enumerate}
\item[2.1)]  $\hat{x}_j,j\in[1,k_1]$ is attacked or
\item[2.2)]  $\hat{v}_j,j\in[1,k_2']$ is attacked or
\item[2.3)]  $\hat{r}_j,j\in[1,k_3]$ is attacked.
\end{enumerate}

For single error on either $(2.1)$ or $(2.2)$, it's not hard to check that node $A$ observes
all $\delta_j\not=0\in\Delta$. Node $A$ can observe received vector,
$$
(\hat{x}_1,\ldots,\hat{x}_{k_1},\hat{v}_1,\ldots,\hat{v}_{k_2'},\hat{r}_1,\ldots,\hat{r}_{k_3})
$$
since $k_3\ge 2$ and $\hat{r}_j$ is redundancy of $\hat{x}_j,\hat{v}_j$, it's not hard to 
see that node $A$ can identify single error symbol by MDS property.

For single error on $(2.3)$, node $A$ observes single $\delta_j\not=0$ of $\Delta$ and
the other $\delta_{j'}=0$, under which node $A$ believes that $\hat{r}_j$ was corrupted.

Once node $A$ identifies error link, it sends $CS$ to node $B$ to trigger node $B$ sending
$W$ to sink.

Case-2-2: $e\ge 2$ in subsequent attack.

For the case that $(2.1)$ or $(2.2)$ happened in first time, assume $\hat{x}_l$ or ${\hat{v}_l}$ was attacked in previous
transmission. Given a received vector $(\ref{received:2})$, node $A$ observes 
\begin{eqnarray}
&& \delta_j-\eta_{l,j}\eta_{l,k_3}^{-1}\delta_{k_3} \nonumber\\
&=& (\sum_{i=1}^{k_1+k_2'}\eta_{i,j}\triangle x_i+\triangle
  r_j)-\eta_{l,j}\eta_{l,k_3}^{-1}(\sum_{i=1}^{k_1+k_2'}\eta_{i,k_3}\triangle x_i+\triangle r_{k_3})
\nonumber  \\
&=&\sum_{i=1}^{k_1+k_2'\backslash l}(\eta_{i,j}-\eta_{l,j}\eta_{l,k_3}^{-1}\eta_{i,k_3})\triangle
x_i+\triangle r_j-\eta_{l,j}\eta_{l,k_3}^{-1}\triangle r_{k_3} \label{observation:1}
\end{eqnarray}
where $\delta_j,\delta_{k_3}\in\Delta,j=1,\ldots,k_3-1$. Since $(\ref{observation:1})$
has removed possible non-zero $\triangle x_l$ detected in first attack, if there is any
non-zero $(\ref{observation:1}),j=1,\ldots,k_3-1$, there must be new error link happened,
node $A$ sends $CS$ signal to node $B$, otherwise node $A$ observes all zero of
$(\ref{observation:1})$, recall that node $A$ has already identified $\hat{x}_l$ or $\hat{v}_l$ in
previous, it sends set $F$ to node $B$, each of $f_j$ defined as
$$
f_j=\left\{
\begin{matrix}
\hat{v}_j+\sum_{i=1}^{k_1}\limits\theta_{i,j}\hat{x}_i\mbox{ and }\sum_{i=1}^{k_1}\limits\theta_{i,l}\hat{x}_i & j\not=l,\mbox{if }\hat{v}_l\mbox{ happened}\\
\hat{v}_j+\sum_{i=1}^{k_1\backslash l}\limits\theta_{i,j}\hat{x}_i &\mbox{if }\hat{x}_l\mbox{ happened}
\end{matrix}
\right .
$$

And similarly as $(\ref{omega:1})$ node $B$ observes 
$$
\omega_j=\left\{
\begin{matrix}
\triangle v_j+\sum_{i=1}^{k_1}\limits\theta_{i,j}\triangle x_i\mbox{ and
}\sum_{i=1}^{k_1}\limits\theta_{i,l}\triangle x_i & j\not=l, \mbox{if }\hat{v}_l\mbox{
  happened}\\
\triangle v_j+\sum_{i=1}^{k_1\backslash l}\limits\triangle x_i &\mbox{if }\hat{x}_l\mbox{ happened}
\end{matrix}
\right .
$$ 
for $f_j\in F,j=1,\ldots,k_2'$.

Notice that both of $(\ref{observation:1})$ and $\omega_j$ has canceled item
$\triangle x_l$ or $\triangle v_l$, together with observation on node $A$, there are $z-1$ equations for $z-1$ at most new injected
error symbols. Since $k_2'+k_3-1=z-1$ and adversary can inject $z-1$ at most error symbols besides $\hat{x}_l$ or
$\hat{v}_l$, if new error position happened, there must be nonzero $\omega_j$, otherwise
it gives contradiction to linear property.

For the case $(2.3)$ happened first time, assume $\hat{r}_l$ was attacked in first time.
Given a received vector $(\ref{received:2})$, node $A$ observes corresponding $\Delta$
first. 

Recall definition of $\Delta$, if there is $\delta_j\not=0,j\not=l$, it indicates nonzero $\triangle x_i,\triangle
r_j,j\not=l$ and must be new error position happened, node $A$ sends $CS$ signal to node $B$. If
node $A$ has observation $\delta_j=0,j=1,\ldots,k_3,j\not=l$, which is
$$
\sum_{i=1}^{k_1+k_2'}\eta_{i,j}\triangle x_{i,j}+\triangle r_j = 0,j\not=l
$$
it sends set $F(\ref{feedback:1})$ to node $B$ and node $B$ follows process $(\ref{omega:1})$ to
observe
$$
\omega_j=\triangle v_j+\sum_{i=1}^{k_1}\theta_{i,j}\triangle x_i,j=1,\ldots,k_2'
$$
If there is new error position happened, there must be nonzero $\omega_j$ which
triggers node $B$ sending claim $W$ to sink, otherwise all $\omega_j=0$ indicate no new error
position happened.

{\bf Case-3}: $|\Delta|=0$, $k'_2=z$

By the condition, since $k_1+k'_2=n$ the row has structure 
\begin{eqnarray}
(x_1,\ldots,x_{k_1},v_1,\ldots,x_{k'_2})=(x_1,\ldots,x_{k_1},x_{k_1+1},\ldots,x_n) \label{received:4}
\end{eqnarray}
where $x_i\in\mathcal{X}$. Let $G$ be generator matrix for $(n+z,n)$ systematic MDS code, and 
$$
G = (I_{n\times n} \;\; E)
$$ 
where $I_{n\times n}$ is unit matrix and $E$ is a $n\times z$ matrix, entry of which is
denoted by $\eta_{i,j},1\le i \le n, 1\le j \le z$, which are chosen uniformly at
random from nonzero entries of a large finite field.

For a received $(\ref{received:4})$, node $A$ operates as
$$
(\hat{x}_1,\ldots,\hat{x}_{k_1},\hat{x}_{k_1+1},\ldots,\hat{x}_{k_1+k'_2})E
$$
and send the following set to node $B$,
\begin{eqnarray}
F = \{ f_j=\sum_{i=1}^n\eta_{i,j}\hat{x}_i | j=1,\ldots,z \} \label{feedback:3}
\end{eqnarray}

Node $B$ receives set $F$, since node $B$ has knowledge about true message symbols, it can
operate as
$$
\Omega=\{\omega_j=f_j-\sum_{i=1}^n\eta_{i,j}x_{i,j}|f_j\in F,j=1,\ldots,z\}
$$
not hard to see that $\omega_j=\sum_i^n\eta_{i,j}\triangle x_i$.

Case-3-1: $e=1$ in first attack.

Node $B$ observes all $\omega_{i,j}\not=0$ in first time, it knows error happened,
node $B$ observes
$$
\frac{\omega_j}{\omega_1} = \frac{\sum_{i=1}^n\eta_{i,j}\triangle
  x_i}{\sum_{i=1}^n\eta_{i,1}\triangle x_i},j=2,\ldots,z
$$
Assume $\hat{x}_l$ was attacked, node $B$ will has observation as
$\frac{\eta_{l,j}}{\eta_{l,1}},j=2,\ldots,z$, thus node $B$ trust that $\hat{x}_l$ was attacked
and sends claim $W$ to sink.

Case-3-2: $e\ge 2$ in subsequent attack.

For this case by applying the proof of Claim~\ref{claim:1}, it will show adversary
cannot hide new error link by attacking multiple links. Thus once there is new error link
happened, it triggers node $B$ sending $W$ to sink.

{\bf Case-4:} $|\Delta|=1,k_2'=z-1$.

By the condition, the row has structure as
\begin{eqnarray}
(x_1,\ldots,x_{k_1},v_1,\ldots,v_{k_2'},r_1) \label{received:3}
\end{eqnarray}

Given a received vector $(\ref{received:3})$, we have
$$
\Delta=\delta_1=\sum_{i=1}^{k_1+k_2'}\eta_{i,1}\triangle x_i+\triangle r_1
$$
The first time node $A$ observes $\delta_1\not=0$, it implies that there
must be nonzero $\triangle x_i,\triangle r_1$, and node $A$ can send $CS$ signal 
to node $B$. However single redundant symbol $r_1$ cannot help node $A$ identifying
erroneous position, it says no matter either $\delta_1=0$ or $\delta_1\not=0$ observed
in subsequent transmission, node $A$ cannot find out if new error position happened.
We need to consider another scheme.  

If certain row has property of $k_2'=z-1$, it indicates that $\lceil\frac{b}{a-c}\rceil\ge z-1$.
And since $b<z(a-c)$ by tight condition, we focus on $z-2<\frac{b}{a-c}<z$.

The intuition of this case is that we first show for most of $b\in((z-2)(a-c),z(a-c))$, we could arrange $k_2'$ for
each row of $X$ properly such that situation of $X$ falls into cases we have solved. Then we discuss
the detection and decoding scheme on more specific condition of $b$.

First we consider network under two conditions: either $b<z(a-c)-1,z>2$ or $b$ is even subjected by $b<z(a-c)-1,
z=2$. It's not hard to check that we could modify encoding scheme slightly such that following 
properties holds for each row of $X$:
\begin{itemize}
\item either $k_2'=z$ for some row of $X$, i.e., $|\Delta|=0$
\item or $0\le k_2'\le z-2$ for some row of $X$, i.e., $2\le |\Delta|\le z$.
\end{itemize}
\begin{IEEEproof}
We first consider the network with $b<z(a-c)-1,z>2$. 
For $b=z(a-c)-\phi, 2\le\phi\le z$, we could assign $k_2'=z$ such that each row of first $(a-c-1)$ rows 
of $X$ carries $z$ $v_i$ symbols giving $|\Delta|=0$ and the last row of $X$ carry $(z-\phi)$
$v_i$ symbols. Notice that
we have $0\le z-\phi \le z-2$ giving $2\le |\Delta|\le z$.

For $b=z(a-c-t)-\phi, 1\le t < a-c, 1\le\phi\le z-1$, we could arrange $k_2'$ such that:
\begin{itemize}
\item Each row of first $(a-c-t-1)$ rows of $X$ carries $z$ $v_i$ symbols, 
\item The $(a-c-t)$th row of $X$ carries $(z-\phi-1)$ $v_i$ symbols,
\item The $(a-c-t+1)$th row carries $1$ $v_i$ symbols,
\item On the remaining row of $X$, no $v_i$ is carried.
\end{itemize}
which gives that $|\Delta|=0$ for first $(a-c-t-1)$ rows of $X$, $2\le |\Delta|\le z$ on
$(a-c-t)$th row of $X$, $|\Delta|=z-1\ge 2$ on $(a-c-t+1)$th row of $X$ and $|\Delta|=z>2$
on the remaining row of $X$.

And for the network with $b$ is even under $b<z(a-c)-1,z=2$, the result is trivial.
\end{IEEEproof}

Detection ability of rows having $|\Delta|=0$ and $2\le |\Delta|\le z$ have been discussed
corresponding to Case-1 and Case-2 respectively and the detection ability of rows having structure
$|\Delta|=z$ will be discussed in Case-4. Therefore we could assume intermediate nodes have
a detectable structure of $X$ under condition either $b<z(a-c)-1,z>2$ or $b$ is even subjected by $b<z(a-c)-1,
z=2$.

Next we consider network under conditions: either $b=z(a-c)-1,z>2$ or $b$ is odd subjected by
$b<z(a-c)-1,z=2$, it's not hard to check that 
\begin{enumerate}
\item[4.1-]\label{fbcondition:1} For $b=z(a-c)-1, z > 2$, we can arrange each row of $k_2'$ such that
$|\Delta|=z$ for each row of first $(a-c-1)$ rows of $X$ and $|\Delta|=1$ for the last row of $X$;
\item[4.2-] $b<z(a-c)-1, z=2$ and $b$ is odd. we could let $X$ has structure as
\begin{itemize}
\item[4.2.1] either $|\Delta|=2$ or $|\Delta|=0$ on the first $(a-c-1)$th row of $X$ and
\item[4.2.2] $|\Delta|=1$ on the last row of $X$.
\end{itemize}
\end{enumerate}

\begin{theorem}
Given $b=z(a-c)-1,z>2$ or $b$ is odd subjected by $b<z(a-c)-1,z=2$, there is a solution 
for sink to decode correctly without claim from node $B$ after sink identified single error on upstream link.
\end{theorem}
\begin{IEEEproof}
Given adversary only corrupts
single link on upstream in first attack and attacks new error links in subsequent transmission. Assume 
that $i$th upstream link was identified in previous transmission, denote the remaining network
by $\mathcal{G}\backslash i$.

{\bf Feedback Condition-1}: $b=z(a-c)-1,z>2$.

Recall $(4.1)$, we first consider rows of received $\hat{X}$ having $|\Delta|=z$ on upstream, 
which is discussed in Case-1, it says that in subsequent transmission if new error link happened,  
intermediate nodes will find it out. If no claim is sent by node $B$, it indicates that the
remaining of these rows are clean. And there are 
totally $(a-c-1)$ rows having $|\Delta|=z$ property.

For the case that there is no new error link happened on these rows,
since it's encoded $n$ independent message symbols in each row and sink has already identified 
one single error link, there is independent symbols as many as,
\begin{eqnarray*}
R_1 = (n-1)(a-c-1)
\end{eqnarray*}
Note that without a little rigorous, we will use $R_1$ to represent these independent symbols on
$(a-c-1)$ rows of $X$ in following.

Then we consider the $(a-c)$th row of received $\hat{X}$ with $|\Delta|=1$. Node $A$ sends received 
$\hat{v}_i,i=1,..,z-1$ to node $B$ in next transmission. Since node $B$ 
knows all true $v_i$ symbols, if there is any symbol corrupted, it must be detected and claim 
will be sent which will let sink has an MDS code, otherwise it does nothing.

With nothing of information from node $B$, it indicates that all these received 
symbols $\hat{v}_i,i=1,..,z-1$ are reliable.

Recall that in the absence of claim from node $B$, sink can trust first $(a-c-1)$ rows of upstream on 
the remaining network $\mathcal{G}\backslash i$. 
Denote the remaining network $\mathcal{G}\backslash i$ excluding first $(a-c-1)$ rows of upstream by
$\mathcal{G}'$. 

We show that there exists a collection of link subset on $\mathcal{G}'$, which has at least $UB-R_1-(z-1)$ 
symbols independent with $R_1,v_i$.

Let link subset on $\mathcal{G}'$ be formulated as: remove links carrying symbols
$\hat{v}_i,i=1,..,z-1$ first, the number of which is $z-1$ and remove another $(z-1)$ 
links as following priority order: first removing upstream
links, and then downstream link until $(z-1)$ links have been removed altogether. 

Not hard to see symbols in the link subset we defined is independent with $R_1,v_i$. Then
we observe the number of symbols in this link subset. We organize discussion by topology of network:

Category-1: $n>2(z-1)$

By above description, the number of independent symbol on the link subset of $\mathcal{G}'$ is
\begin{eqnarray}
R_2 = (n-1-2(z-1))(c+1)+mc
\end{eqnarray}
Then we see,
\begin{eqnarray}
&& R_1+R_2+(z-1)  \nonumber \\
&=& (n-z)a+(m-z)c+(z-1)a-(z-2)c-z \nonumber \\
&=& UB-b+(z-1)a-(z-2)c-z \nonumber \\
&\ge& UB+2c-a-z+2 \label{eqn:bb1}
\end{eqnarray}
where $(\ref{eqn:bb1})$ follows that $b\le z(a-c)-1$ and $(z-1)$ is from the number 
of $\hat{v}_{i},i=1,...,z-1$. 

Since $b<\min\{z(a-c),zc-(z-2)a\}$ by tight condition $(\ref{con1})$, we focus on the 
condition $z(a-c)\le zc-(z-2)a$, otherwise we cannot have $b=z(a-c)-1$. 
Then we have 
\begin{eqnarray}
\frac{c}{a} &\ge& \frac{z-1}{z} \label{ineq:2} \\
\frac{a}{a-c}&\ge& z \nonumber
\end{eqnarray}
Note that we don't have to consider tight condition $(\ref{con3})$ since $(m-z)(a-c)<z(a-c)$
for $m<2z$, no matter how there is always $b<z(a-c)-1$.

Then we observe,
\begin{eqnarray} 
2c-a-z+2 &\ge& 2c-a-\frac{a}{a-c}+2 \label{ineq:4} 
\end{eqnarray}
where $(\ref{ineq:4})$ follows $(\ref{ineq:2})$, let $a-c=x$, we have
\begin{eqnarray}
&& 2c-a+2-\frac{a}{x}=\frac{-2x^2+(a+2)x-a}{x} \nonumber 
\end{eqnarray}
not hard to see that the root of $-2x^2+(a+2)x-a$ are $x_1=1,x_2=\frac{a}{2}$,
which says that as long as $1 \le x \le \frac{a}{2}$, the result of $-2x^2+(a+2)x-a$ are 
no less than zero.

Since $x=a-c$, recall $(\ref{ineq:2})$ and $z\ge 2$, not hard to check that we 
have $1\le x \le \frac{a}{2}$, which shows that
\begin{eqnarray}
&& 2c-a+2-\frac{a}{x} \nonumber \\
&=& \frac{-2x^2+(a+2)x-a}{x} \ge 0 \nonumber 
\end{eqnarray}
Then we have 
\begin{eqnarray}
R_1+R_2+(z-1) \ge UB \label{ineq:5}
\end{eqnarray}

Category-2: $z< n \le 2(z-1)$

Similarly, on the link subset of $\mathcal{G}'$, the independent symbol is 
\begin{eqnarray}
R_2 &=& (m-(2(z-1)-(n-1)))c \nonumber \\
   &=& (n+m-2z+1)c \nonumber
\end{eqnarray}
then the total reliable symbols we have
\begin{eqnarray}
&& R_1+R_2+(z-1) \nonumber \\
&=& (n+m-2z+1)c+(n-1)(a-c-1)+(z-1) \nonumber \\
&=& UB-b+(z-1)a-(z-2)c-(n-z) \nonumber \\
&\ge& UB+2c-a-(n-z)+1  \label{eqn:bb2} \\
&=& UB+2c-a-n+z+1 
\end{eqnarray}
where $(\ref{eqn:bb2})$ follows that $b\le z(a-c)-1$. 

And since tight condition $(\ref{con2})$, i.e., $b<\min\{z(a-c),(n-z+2)c-(n-z)a\}$, 
we consider the case $z(a-c)\le (n-z+2)c-(n-z)a$, which gives 
\begin{eqnarray}
\frac{c}{a-c} &\ge& \frac{n}{2} \nonumber \\
\frac{2a}{a-c}-2 &\ge& n \label{ineq:6}
\end{eqnarray}
also we have $z(a-c)\le zc-(n-z)a$ since $n\le 2(z-1)$, which gives 
$z\ge \frac{na}{2c}$. Similarly, we ignore tight condition $(\ref{con4})$.

Then we observe,
\begin{eqnarray*}
&& 2c-a-n+z+1  \\
&\ge& 2c-a-n+\frac{na}{2c}+1  \\
&=& 2c-a-(1-\frac{a}{2c})n+1  \\
&\ge&2c-a-\frac{2c-a}{2c}(\frac{2a}{a-c}-2)+1  \\
&=&2c-a-\frac{2c-a}{2c}\frac{2c}{a-c}+1  \\
&=&2c-a-\frac{a}{a-c}+\frac{2(a-c)}{a-c}+1  \\
&=& 2c-a-\frac{a}{a-c}+3 
\end{eqnarray*}
In category-1, we have shown that $2c-a-\frac{a}{a-c}+2\ge 0$, thus in category-2, 
we have 
\begin{eqnarray}
R_1+R_2+(z-1)>UB \label{ineq:7}
\end{eqnarray}

Notice that above proof all shows that network we need to consider is subjected
by $m\ge 2z$.

{\bf Feedback Condition-2}: $b$ is odd subjected by $b<z(a-c)-1,z=2$.

Recall $(4.2)$, similarly we first consider rows having either $|\Delta|=z$ or $|\Delta|=0$ of
received $\hat{X}$. By Case-1 and Case-3, we know intermediate nodes can detect new error position
after first attack. Also there $(a-c-1)$ rows satisfying these condition.

Then we observe the independent symbols for the case that no new error link happened after first
attack. Since there are $\lfloor\frac{b}{2}\rfloor$ rows having $|\Delta|=z$ and there are
$b-1$ $v_i$ symbols on first $(a-c-1)$th rows, so we should have
$$
R_1=(n-2)(a-c-1)+(b-1)-\lfloor\frac{b}{2}\rfloor
$$
where $\lfloor\frac{b}{2}\rfloor$ is number of symbols lost on identified single link.

Following the same logic in Feedback Condition-1, 

Category-1: $n>2(z-1)$,
we have
$$
R_2 =(n-1-2(z -1))(c+1)+mc=(n-3)(c+1)+mc
$$
Then we see
\begin{eqnarray*}
R_1+R_2+1 &=& UB+c-\lfloor\frac{b}{2}\rfloor-1  \\
         &>& UB+c-1-\lfloor\frac{2(a-c)-1}{2}\rfloor  \\
         &>& UB+c-1-(a-c-1)  \\
         &=&UB+2c-a 
\end{eqnarray*}
recall $(\ref{ineq:2})$, we have $\frac{c}{a}\ge\frac{1}{2}$, which indicates
that
\begin{eqnarray}
R_1+R_2+1\ge UB \label{ineq:9}
\end{eqnarray}

Category-2: $z<n\le 2(z-1)$. Obviously, there is no such network.

Next we show by the result $(\ref{ineq:5}),(\ref{ineq:7}),(\ref{ineq:9})$, we can design a decoder for sink. Specifically, 
let set $S$ be all forward link on $\mathcal{G}'$. Define two subsets of $S$ as,
\begin{eqnarray}
S_1 &=& \{ \mbox{  links which carry symbols } v_i \mbox{ on } \mathcal{G}'\} \nonumber \\
S_2 &=& S\backslash \{i,S_1\} \nonumber
\end{eqnarray}
where $i$ is the index of link identified in previous transmission. 

The intuition of decoder is to directly consider different sets of $z-1$ links
as removing set and check the remaining links for consistency. When it finds a 
consistent set, the decoding process ends. Thus we consider all possible sets
of $2(z-1)$ links and show that the remaining links carry a full rank complete
decoding set.

The remaining symbols must contain a full rank non-erroneous set from which all the
message symbols can be correctly decoded, along with others
that may or may not be erroneous. Therefore if the considered
subset (removing set) includes all the actual erroneous links, the remaining
symbols will be consistent and give the correct decoding,
whereas if the remaining symbols contain errors there will
be inconsistency with the full rank non-erroneous set.

The decoder exhaustively goes through all sets of $(z-1)$ forward links in turn,
checking if the remaining forward links are consistent. If they are consistent, they
give the correct message. Let new error link set be $E=E_1\cup E_2$ in $\mathcal{G}'$, 
where $E_1\subseteq S_1$, $E_2\subset S_2$ and $|E_1|+|E_2|\le z-1$.

For the case that $E=E_1\subseteq S_1$, i.e., all new $(z-1)$ at most error happened 
in $S_1$, note that since $\hat{v}_i$ can be detected on node $B$, we consider the 
new errors that occurred on $Y$ of $S_1$ for this case. If we consider the remaining link subset 
by $(\ref{ineq:5}),(\ref{ineq:7}),(\ref{ineq:9})$, 
it indicates that under $E=E_1$ it will give more than one full rank non-erroneous
link subset since $|S_2|>z-1$.

Starting from this case, we show decoder can finish decoding process as following order:
\begin{enumerate}
\item[O-1] Assume $E=E_1\subseteq S_1$, i.e., all new errors happened in $S_1$, observing 
whether there is consistent result by removing $\{S_1,\tilde{S}\}$, where $\tilde{S}\subset 
S_2$ is followed by priority order. 
If it successes, output the result, otherwise goes to next,
\item[O-2] Assume both of $E_1$ and $E_2$ are non-empty, then make consistent check in remaining link
subset. If it successes, output the result, otherwise goes to next,
\item[O-3] Above two processes failed, it indicates that $E=E_2\subset S_2$, decoder decode in
remaining links.
\end{enumerate}

O-1: New errors set $E$ only happened in $S_1$, i.e., $E=E_1\subseteq S_1$.

Decoder first removes links in $S_r=\{S_1,\tilde{S}\}$,
where  $\tilde{S}\subset S_2$ and $|\tilde{S}|=z-1$. Since $|S_2|>z-1$, we can have more than one 
subset $\tilde{S}$. Recall $(\ref{ineq:5}),(\ref{ineq:7}),(\ref{ineq:9})$,  the remaining link subset together with
$R_1,v_i$ will provide no less than $UB$ non-erroneous symbols. And clearly this kind of link subset
is more than one.

Clearly for $E\subseteq S_1$, correct decoding occurs since all possible remaining link subsets together with $R_1,v_i$
will give consistent result. However, if $E\not\subseteq S_1$, it indicates that by changing
$\tilde{S}$, there must be an inconsistent result, then decoder goes to next.

O-2: New errors happened in both of $S_1$ and $S_2$, i.e., $|E_1|\not=0,|E_2|\not=0,
0<|E_1|+|E_2|\le z-1$. We assume $|E_1|\not=0$, it implies that $z>2$, otherwise
$|E_2|=0$ for $z=2$ since one error link has been identified.

O-1 decoded unsuccessfuly, it implies that $S_1$ doesn't include all
new error links, i.e., $|E_2|\not=0$.

Consider removing set $S_r=\{S_1,\tilde{S}\}$, $S_1$ includes $E_1$ erroneous subset, if
$E_2\subset \tilde{S}$, similarly, by $(\ref{ineq:5}),(\ref{ineq:7}),(\ref{ineq:9})$, 
the sum of $R_1,S\backslash S_r$ and $\hat{v}_i,
$ is no less than $UB$, and we show one consistent result at least will be given.

Let $\tilde{S}=\{E_2,S_3\}$, where $S_3\subset S_2\backslash E_2,
|S_3|+|E_2|=z-1$. Since $|E_2|<z-1$, it implies that $|S_3|\ge 1$. Note that if we hold $E_2$
unchanged and change set $S_3$, we could have a group of $\tilde{S}$ which have different subset $S_3$.

Recall that $E$ is new error link set, let $\triangle E$ is injected error vector set corresponding
to $E$, the element of which is injected error vector denoted by $\triangle e_i$ on $e_i\in E$.

For an set $S_r$, suppose $S\backslash S_r$ includes error link, it's
not hard to see that the result from $(S\backslash S_r,R_1,v_i)$ will pollute the
original clean message symbol, we claim:
\begin{claim}\label{errorclaim}
The polluted message symbol resulted from $(S\backslash S_r,R_1,v_i)$ equals to the clean symbol
injected by a linear combination of erroneous in $\triangle E$.
\end{claim}
\begin{proof}
Consider we compute message symbols from $(S\backslash S_r,R_1,v_i)$.
Suppose $S\backslash S_r$ includes erroneous, let $\hat{c}=c+\triangle e$ be polluted symbols
in $S\backslash S_r$, where $c$ is clean symbol and $\triangle e$ is injected error.

According to encoding scheme, each symbol in $(S\backslash S_r,R_1,v_i)$
can be written,
\begin{eqnarray*}
\hat{c} &=& H({\bf x},{\bf y}) \\
        &=& c+\triangle e
\end{eqnarray*}
where ${\bf x},{\bf y}$ represent for message symbols $x\in\mathcal{X},y\in\mathcal{Y},
$ respectively and $H(\cdot)$ is a linear function
determined by encoding scheme. 

Since $|{\bf x}|+|{\bf y}|=UB$ and $|(S\backslash S_r,R_1,v_i)|\ge
UB$, we have enough independent linear function $H({\bf x},{\bf y})$ to compute
${\bf x},{\bf y}$. And by the property of linear, it's not hard to see that the result
of $\hat{{\bf x}},\hat{{\bf y}}$ equals to 
\begin{eqnarray*}
\hat{{\bf x}} &=& {\bf x}+H_1(E) \\
\hat{{\bf y}} &=& {\bf y}+H_2(E) 
\end{eqnarray*}
where $H_i(E)$ is a linear function for injected error
set determined by encoding scheme.
\end{proof}

Then we see the following claim.
\begin{claim}\label{inconsistentclaim}
There exists  inconsistent output for two removing sets $\tilde{S}^1=\{E'_2,S^1_3\},
\tilde{S}^2=\{E'_2,S^2_3\}$ where $E'_2$ is subset of $E_2$, $|E'_2|<|E_2|$, 
and $S_3^i$ is error free subset of $S_2$, $|E'_2|+|S_3^i|=z-1,i=1,2$.
\end{claim}
\begin{proof}
We consider the case that $|E_1|=1, |E_2|=z-2$ and $|E'_2|=0$, which says the removing subset
$\tilde{S}^i,i=1,2$ doesn't includes any error link, i.e., $E_2\in S_2\backslash \tilde{S}^i$, giving
$\tilde{S}^1=S_3^1,\tilde{S}^2=S_3^2$. This is obviously the worst case since $S_2\backslash
\tilde{S}^i$ includes all $z-2$ error links.

Let $S_3^1=\{S',i\}, S_3^2=\{S',j\}$, where $i,j,i\not=j$ are index of forward link and 
$i,j\not\in S',|S'|=z-2$. Then we use $S_3^1(i),S_3^2(j)$ to represent for $\tilde{S}^1,\tilde{S}^2$ 
respectively.

Assume $S\backslash \{S_1,S_3^1(i)\},S\backslash \{S_1,S_3^2(j)\}$ give consistent 
result together with $R_1,v_i$ respectively. Note that according \\
to result from $(S\backslash S_r,R_1,v_i)$, where $S_r\in\{\{S_1,S_3^1(i)\},\{S_1,\\
S_3^2(j)\}\}$
, sink can compute the symbols which are in the removing set $S_3^1(i),S_3^2(j)$. 
Then since the consistency occurred, we should have following properties between $S_3^1(i)$
and $S_3^2(j)$:

Symbols on $j$th link computed from the result of $(S\backslash \{S_1,S_3^1(i)\},R_1,v_i)$ 
are consistent with symbols on $j$th in $S_3^2(j)$.

Define $c_{j,i}$ is a symbol on $j$th link computed from the result of $(S\backslash
\{S_1,S_3^1(i)\},R_1,v_i)$ and $c_{i,j}$ is a 
symbol on $i$th link computed from the result of $(S\backslash
\{S_1,S_3^2(j)\},R_1,v_i)$. Similar, we use $c_{i,i}$ as a symbol 
on $i$th link in $S_3^1(i)$ and $c_{j,j}$ as a symbol on $j$th link in $S_3^2(j)$. Since $S_3^1(i)$
and $S_3^2(j)$ are error free link set by definition, $c_{i,i}$ and $c_{j,j}$ are true symbols, by
above, we have,
\begin{eqnarray}
c_{j,i} &=& c_{j,j}+H_i({\bf e}) \nonumber \\
       &=& c_{j,j} \nonumber 
\end{eqnarray}
where $H_i(\cdot)$ is linear function of injected error determined by coding scheme,
which ${\bf e}$ is linear combination of error vector injected in $z-2$ links by Claim
\ref{errorclaim} and belongs to $S\backslash
\{S_1,\tilde{S}^i\},i=1,2$. Since consistency occurred, we have $c_{j,i}=c_{j,j}$, which 
implies that $H_i({\bf e})=0$. 

Suppose that $i$ and $j$ in $S_3^1(i),S_3^2(j)$ respectively are the index of downstream links, 
it implies that there are $c$ equations having $H_j({\bf e})=0,j=1,...,c$.

Then decoder changes $j$ in $S_3^2(j)$ to $k$, where $k\not=j\not=i,k\not\in S'$ and
let $k$ be index of downstream. Repeat above observation, since consistency happened,
it gives another group of linear functions $H_k({\bf e})=0,k=1,..,c$. And note that $m\ge 2z$
for $z>2$, thus we have at least $(m-1)c$ linear function $H_i({\bf e})=0$.

Recall that we have $(z-2)a$ injected error at most in $E_2$, by $(\ref{ineq:2}),(\ref{ineq:6})$, 
it's not hard to see that $2c\ge a$, which gives $(m-1)c\ge (z-2)a$, and totally we have $H_i({\bf e})=0,i=1,..,(m-1)c$  
, which indicates that ${\bf e}$ equals to zero,
it obviously contradicts to our assumption that we should have non-zero injected error vector ${\bf e}$.
\end{proof}

Claim \ref{inconsistentclaim} indicates a decoding strategy: there is only one removing set 
$S_r=\{S_1,\tilde{S}\}$, where $\tilde{S}=\{E_2,S_3\}, |E_2|<z-1, |S_3|>0, |E_2|+|S_3|=z-1, 
S_3\in S_2$ such that when holding $E_2$ unchanged, sink changes $S_3$, the $(S\backslash S_r, 
R_1, v_i)$ give the consistent output.

If sink cannot find $E_2$ satisfying above requirement, it says assumption 3 happened and goes to
next case.

{\bf Case-3}: New error only happened in $S_2$ and $E=E_2\subset S_2$, i.e., $|E_1|=0,|E_2|=z-1$.

In this case the symbols in $S_1$ are clean. Still we consider removing set
$S_r=\{S_1,\tilde{S}\}$ first. Sink can have a result from $(R_1,S\backslash S_r,v_i)$.

Since $S_1$ is clean under this case, sink could trust symbols in it. Then sink computes
$v_i$ by the result from $(S\backslash S_r,R_1,v_i)$. If the 
result is inconsistent with corresponding symbols in $S_1$, sink choose different $S_r$ 
to repeat above procedure till consistency occurs. We show that sink can trust this result.
By occurred consistency, it gives,
\begin{eqnarray}
v_{i} &=& H_i(\hat{v}_i) \nonumber \\
    &=& H_i(v_i)+H_i(\triangle v_i) \nonumber 
\end{eqnarray}
where $v_i\in S_1$ and $\hat{v_i}$ are symbols computed from the result of 
$(S\backslash S_r,R_1,v_i)$, $\triangle v_i$ are
linear combination of injected error vector by Claim \ref{errorclaim} and $H_i(\cdot)$ 
is a linear function determined by encoding scheme.

Since consistency,  $H_i(\triangle v_i)=0$. Similarly, we will have,
\begin{eqnarray*}
y_j &=& H_j(\hat{{\bf y}}) \\
    &=& H_j({\bf y})+H_j(\triangle {\bf y})
\end{eqnarray*}
where $y_j\in S_1$ and $\hat{{\bf y}}$ are symbols computed from the result of $(S\backslash
S_r,R_1,v_i)$, $\hat{{\bf y}}={\bf y}+\triangle {\bf y}$, 
${\bf y}$ represent true symbols, $\triangle {\bf y}$ are linear combination of injected error vector 
by Claim \ref{errorclaim}  and $H_j(\cdot)$ is a linear function determined by encoding scheme.
Then we have $H_j(\triangle {\bf y})=0$.

Note that each $\triangle {v_i},\triangle {\bf y}$ is linear combination of
injected error vector and we have $(z-1)$ links at most corrupted , since $|S_1|=z-1$, 
with all $H_j(\triangle {\bf y})=0,H_i(\triangle v_i)=0$, it indicates that
these all injected error equals to zero. Thus sink can trust the result from $(S\backslash S_r,
R_1,v_i)$.
\end{IEEEproof}

\subsection{Summary of Error Signaling Procedure} 
For received $\hat{X}$, we define $\Delta_1, l=1,\ldots, a-c$ for each row of
$\hat{X}$ as definition $(\ref{Delta2})$. We summarize the error signaling procedure 
on intermediate node $A$ and $B$ for as follows:
\begin{enumerate}
\item For $|\Delta_l|=z$, node $A$ observes nonzero $\delta_j\in\Delta_l$ first time, it sends
$CS$ single to node $B$ and node $B$ sends $W$ to sink. And Claim~1 shows adversary cannot
hide new error position for node $A$ in subsequent attack.
\item For $2\le|\Delta_l|<z$, if node $A$ observes nonzero $\delta_j\Delta_l$ first time it
sends $CS$ to node $B$; for observation all $\delta_j=0$, it sends set $(\ref{feedback:1})$ to 
node $B$, either of which will help intermediate nodes identifying happened error.

In subsequent transmission, node $A$ and $B$ cooperate following Case~2-2.

\item For $|\Delta_l|=0$, each time node $A$ sends set $(\ref{feedback:3})$ to node $B$,
according which node $B$ can identifying new error position once it happened.
\item For $|\Delta_l|=1$, intermediates nodes cooperates directly as Case 4 described.
\end{enumerate}
Notice that for all rows of $X$, if no new error position is detected, amount of transmission
on feedback equals to $b$.

\subsection{Summary}
By above analysis, when node $A$ sends $CS$ signal or node $B$ sends claim to sink, it
causes overhead problem on feedback and downstream, however
this overhead only happens when there is new error link occured, and by
Lemma~\ref{Singleton-new} after identifying two error links, sink doesn't need any
side information from feedback or claim. Once either node $A$ sends two times $CS$ signal or node
$B$ sends claim two times, it doesn't need any side information in future
transmission, which means that comparing to the whole transmission, these overhead can be ignored.

\section{Decoding}
\label{decoding}

At steps in which error signaling occurs, all information is forwarded to the sink, 
enabling it to identify the erroneous links and decode correctly as described in the 
previous subsection. In this section we show that the sink is also able to decode correctly 
at steps in which no error signaling occurs. Let $e$ be the number of previously identified 
erroneous links. The sink considers the network excluding these $e$ links, which can contain 
at most $z-e$ erroneous links.

The sink considers each subset of $z-e$ links in turn, assuming it to be the erroneous set 
and checking whether this assumption is consistent with its observations, i.e.~whether the 
symbols that are assumed to be non-erroneous have consistent values. We will show that if 
the remaining symbols are consistent, then they give the correct decoding.  The argument 
shows that the equations corresponding to the remaining symbols must contain a full rank 
non-erroneous set from which all the message symbols can be correctly decoded, along with 
others that  may or may not be erroneous. Therefore if the considered subset includes all 
the actual erroneous links, the remaining symbols will be consistent and give the correct 
decoding, whereas if the remaining symbols contain errors there will be inconsistency with 
the full rank non-erroneous set. The arguments relies on the property that except for 
redundant symbols on the upstream links, the remaining code symbols received by the sink 
form an MDS code.

For the case $z=1$, if an error has previously been signaled, the single erroneous link has
been identified. If no errors have been signaled, this means that there are no errors in $X$.
Therefore, message symbols $\mathcal{X,F}$ are decoded correctly from $X$, and the MDS property
of the generic linear code ensures correct decoding of message symbols $\mathcal{Y}$ under
errors in $Y$ on any single link.

We discuss the case $z\ge 2$ as following structure:

Case 1: $e\ge 2$.

In this case, by Lemma 1 the feedback link is no longer needed. We show that removing any $ 2(z-e)$ links 
in addition to the $e$ identified erroneous links still leaves a full rank non-erroneous set  
along with at least one additional symbol.
To find the worst-case number of remaining non-erroneous MDS symbols at the sink, we  remove 
$2z-e$ links from the original network in the following priority order: first removing upstream 
links, and then downstream links, until $2z-e$ links have been removed altogether, with $e$ set to 
its minimum value, $2$. Note that $2z-2\ge z$ since $z\ge 2$.

Case 1-1: $2z-2\le n$. The worst-case number of remaining non-erroneous MDS symbols is $K_1
=(a-c)(n+2-2z)+c(n+m+2-2z))=a(n+2-2z)+cm$. 
For category 1 and 3 networks, applying the inequality $b \le zc-(z-2)a$ from (\ref{con1})
and (\ref{con3}), 
we have
\begin{eqnarray*}
K_1 -UB &=& (n+2-2z)a+mc\\
       & &-((n-z)a+(m-z)c+b)  \\
       &=& 2a-z(a-c)-b \\
       &\ge& 0.
\end{eqnarray*}Thus,  the condition for correct decoding is satisfied. For category 2 and 4
networks, applying the inequality $b \le (n-z+2)c-(n-z)a$ from (\ref{con2}) and (\ref{con4}), we have
\begin{eqnarray*}K_1 -UB
     &=& 2a-z(a-c)-b \\
     & \ge &a( 2-2z +n)\\& \ge& 0
\end{eqnarray*}
where the last inequality follows from the assumption $2z-2\le n$. 
This also satisfies the condition for correct decoding.

Case  2a-2: $2z-2> n$.  In this case, the network falls in category 2 or 4, 
and the worst-case number of remaining non-erroneous MDS symbols is $K_2=c(m+n-2z+2)$. 
Applying the inequality $b \le (n-z+2)c-(n-z)a$ from (\ref{con2}) and (\ref{con4}), we have
\begin{eqnarray*}
K_2 -UB & = &2c-(n-z)(a-c)\\& &-(n-z+2)c-(n-z)a\\& \ge &0.
\end{eqnarray*}
Thus, the condition for correct decoding is satisfied.

Case 2: $e<2$ links not including the feedback link have previously been identified 
as erroneous. By the arguments in the previous subsection, if no error is being signaled, 
there are two possibilities: either there are no errors in $X$ on the remaining links, 
or the feedback link is erroneous.

Case 2-1: The feedback link is erroneous.  For any size-$(z-e)$ subset considered by the sink, 
at most $z-e-1$ other forward links in the remaining network are erroneous.

If the size-$(z-e)$ subset considered by the sink contains the feedback link, the worst-case 
number of remaining non-erroneous MDS symbols is obtained by removing $2z-e-2$ forward links 
from the original network in the same priority order as in Case 1, with $e = 0$. This is equal 
to $SB2$, which is greater than or equal to $UB$ by the conditions (\ref{con1}-\ref{con4}), 
so correct decoding occurs.

If the size-$(z-e)$ subset considered by the sink does not contain the feedback link, 
the corresponding assumption is that all the  symbols in $X$ other than those on the $e$ 
identified erroneous links are correct. The worst-case number of remaining non-erroneous MDS symbols 
is obtained by removing $z-1$ forward links from the original network in the same priority order as
in Case 1,  along with the symbols in $Y$ from $z-e$ other links, with $e=0$,
This gives $a(n-z)+c(m-z)+a$ remaining non-erroneous
MDS symbols, which is greater than $UB$, so correct decoding occurs.

Case 2-2: The feedback link is not erroneous. In this case, there are no errors in $X$ 
on the remaining links.

If the size-$(z-e)$ subset considered by the sink contains the feedback link, the worst-case number
of remaining non-erroneous MDS symbols is obtained by removing $z-1$ forward links from the original 
network in the same priority order as in Case 1,  along with the symbols in $Y$ from $z-e$ other
links, with $e= 0$. 
This gives $a(n-z)+c(m-z+1)+k$ remaining non-erroneous MDS symbols, which is greater than 
$UB$, so correct decoding occurs.

If the size-$(z-e)$ subset considered by the sink does not contain the feedback link, 
the corresponding assumption is that all the symbols in $X$ other than those on the $e$ 
identified erroneous links are correct. The worst-case number of remaining non-erroneous MDS 
symbols is obtained by removing  the symbols in $V$ from $2z-e$ links, with $u = 0$, and noting 
that $z(a-c)-b$ of the remaining symbols are not MDS. This gives $a(n-z)+c(m-z) +b$ remaining 
non-erroneous MDS symbols, which is equal to $UB$, so correct decoding occurs.

\section*{acknowledgements}
This work is partly supported by National Natural Science Foundation of China (Ref.No. 60832001 \& 
Ref.No. 60672119), the Air Force Office of Scientific Research under grant FA9550-10-1-0166 and Caltech's
Lee Center for Advanced Networking.

\bibliographystyle{IEEEtran}
\bibliography{Yanbo_fb}

\begin{thebibliography}{1}
\providecommand{\url}[1]{#1}
\csname url@samestyle\endcsname
\providecommand{\newblock}{\relax}
\providecommand{\bibinfo}[2]{#2}
\providecommand{\BIBentrySTDinterwordspacing}{\spaceskip=0pt\relax}
\providecommand{\BIBentryALTinterwordstretchfactor}{4}
\providecommand{\BIBentryALTinterwordspacing}{\spaceskip=\fontdimen2\font plus
\BIBentryALTinterwordstretchfactor\fontdimen3\font minus
  \fontdimen4\font\relax}
\providecommand{\BIBforeignlanguage}[2]{{%
\expandafter\ifx\csname l@#1\endcsname\relax
\typeout{** WARNING: IEEEtran.bst: No hyphenation pattern has been}%
\typeout{** loaded for the language `#1'. Using the pattern for}%
\typeout{** the default language instead.}%
\else
\language=\csname l@#1\endcsname
\fi
#2}}
\providecommand{\BIBdecl}{\relax}
\BIBdecl

\bibitem{kim2010necjournal}
S.~Kim, T.~Ho, M.~Effros, and S.~Avestimehr, ``{Network error correction with
  unequal link capacities},'' \emph{IEEE Transactions on Information Theory},
  2011, to appear.

\bibitem{cai2006network}
N.~Cai and R.~Yeung, ``{Network error correction, part II: Lower bounds},''
  \emph{Communications in Information and Systems}, vol.~6, no.~1, pp. 37--54,
  2006.

\bibitem{yeung2006network}
R.~Yeung and N.~Cai, ``{Network error correction, part I: Basic concepts and
  upper bounds},'' \emph{Communications in Information and Systems}, vol.~6,
  no.~1, pp. 19--36, 2006.

\bibitem{kim2009network}
S.~Kim, T.~Ho, M.~Effros, and S.~Avestimehr, ``{Network error correction with
  unequal link capacities},'' in \emph{47th Annual Allerton Conference on
  Communication, Control, and Computing}, 2009.

\bibitem{kim2010nec}
------, ``{New results on network error correction: capacities and upper
  bounds},'' in \emph{Information Theory and Applications Workshop (ITA),
  2010}.\hskip 1em plus 0.5em minus 0.4em\relax IEEE, 2010, pp. 1--10.

\bibitem{kosut2009nonlinear}
O.~Kosut, L.~Tong, and D.~Tse, ``{Nonlinear network coding is necessary to
  combat general byzantine attacks},'' in \emph{47th Annual Allerton Conference
  on Communication, Control, and Computing}, 2009.

\bibitem{kosut-polytope}
------, ``{Polytope codes against adversaries in networks},'' in
  \emph{Information Theory Proceedings (ISIT), 2010 IEEE International
  Symposium on}.\hskip 1em plus 0.5em minus 0.4em\relax IEEE, 2010, pp.
  2423--2427.

\bibitem{Feedbackita}
T.~Ho, S.~Kim, Y.~Yang, M.~Effros, and S.~Avestimehr, ``{On network error
  correction with limited feedback capacity},'' in \emph{Information Theory and
  Applications Workshop (ITA), 2011}.\hskip 1em plus 0.5em minus 0.4em\relax
  IEEE, 2011.

\bibitem{Feedbackyanbo}
Y.~Yanbo and H.~Tracey, ``{Network error correction with limited feedback
  capacity}.''

\end{thebibliography}

\end{document}